# Low-frequency modes in the Raman spectrum of *sp-sp$^2$* nanostructured carbon

C.S. Casari[1], A. Li Bassi[1], A. Baserga[1], L. Ravagnan[2], P. Piseri[2], C. Lenardi[3],
M. Tommasini[1], A. Milani[1], D. Fazzi[1], C.E. Bottani[1], P. Milani[2]

[1] Dipartimento di Chimica Materiali e Ingegneria Chimica 'G. Natta' and
NEMAS – Center for NanoEngineered Materials and Surfaces
Politecnico di Milano, P.zza Leonardo da Vinci 32 I-20133 Milano, Italy

[2] Dipartimento di Fisica
CIMAINA – Interdisciplinary Centre for Nanostructured Materials and Interfaces
Università di Milano, Via Celoria 16 I-20133 Milano, Italy

[3] Istituto di Fisiologia Genenerale e Chimica Biologica
CIMAINA – Interdisciplinary Centre for Nanostructured Materials and Interfaces
Università di Milano, Via Trentacoste 2, I-20134 Milano, Italy

**Abstract**

A novel form of amorphous carbon with *sp-sp$^2$* hybridization has been recently produced by supersonic cluster beam deposition showing the presence in the film of both polyynic and cumulenic species [L. Ravagnan et al. Phys. Rev. Lett. 98, 216103 (2007)]. Here we present a in situ Raman characterization of the low frequency vibrational region (400-800 cm$^{-1}$) of *sp-sp$^2$* films at different temperatures. We report the presence of two peaks at 450 cm$^{-1}$ and 720 cm$^{-1}$. The lower frequency peak shows an evolution with the variation of the *sp* content and it can be attributed, with the support of density functional theory (DFT) simulations, to bending modes of *sp* linear structures. The peak at 720 cm$^{-1}$ does not vary with the *sp* content and it can be attributed to a feature in the vibrational density of states activated by the disorder of the *sp$^2$* phase.

## I. Introduction

1D carbon nanostructures with *sp* hybridization are the subject of an increasing interest motivated by their physical properties and in view of technological applications [1-3]. *sp* carbon chains have been extensively studied both theoretically [4-8] and experimentally as isolated species in the gas phase, in solutions, in solid inert matrices and in carbon nanotubes [9-15]. Pure carbon solids containing macroscopic fraction of *sp*-coordinated atoms have been predicted theoretically [16-18] since decades but only recently their existence has been confirmed experimentally [19,20]. Generally, *sp* coordination is expected to be characteristic of small carbon clusters (less than a few tens of atoms) with linear or ring structure [21-23], much less is known about solid state *sp*-rich carbon systems since the stability of the *sp* phase compared to $sp^2$ and $sp^3$ hybridization is still an open issue [1, 24,25].

We have recently demonstrated the production, by supersonic cluster beam deposition, of nanostructured carbon films with *sp*-$sp^2$ hybridization where polyynic and cumulenic species are present [19]. Structural and vibrational properties of *sp*-$sp^2$ amorphous carbon have been characterized by Raman spectroscopy [20, 24]: G and D Raman bands between 900 and 1700 $cm^{-1}$ typical of $sp^2$ amorphous carbons coexist with features in the 1800-2200 $cm^{-1}$ region which are the fingerprints of the presence of *sp* coordinated carbon atoms since they originate from stretching modes of C-C bonds [6,10].

The vibrational frequency region below 900 $cm^{-1}$ of *sp*-$sp^2$ carbon has not yet been explored and even for $sp^2$ amorphous carbons there is no general agreement about the origin of the features observed at about 400-500 $cm^{-1}$ and 700-800 $cm^{-1}$ in several amorphous carbon-based systems and generically attributed to peaks activated by the structural disorder of the $sp^2$ phase through the relaxation of the q=0 selection rule [26-36]. At present low frequency modes still lack for a detailed comprehension.

Here we report an *in situ* Raman spectroscopy characterization of the low-frequency region of *sp*-$sp^2$ cluster-assembled carbon films showing the presence of two features peaked at 450 and 720 $cm^{-1}$. By varying the film substrate temperature during and after cluster deposition we could vary the *sp* content and correlate this variation to the behaviour of the two peaks. The results are interpreted in the light of DFT numerical simulations of finite and infinite *sp* chains allowing to highlight the contribution of *sp* and $sp^2$ hybridizations to the observed vibrational features.

## II. Experimental Details

Nanostructured carbon films were grown in UHV conditions by depositing supersonic cluster beams produced by a pulsed microplasma cluster source (PMCS) (as described in more detail in [37-39]). The cluster mass distribution, measured by a time-of-flight mass spectrometer prior to deposition [40] is log-normal and it is peaked at about 600 atoms/cluster, extending up to several thousands atoms/cluster. Mass spectra are characterized by the presence of clusters with even and odd number of atoms with no magic numbers in the mass abundance corresponding to fullerene structures [41,42]. Clusters are accelerated by the supersonic expansion at a kinetic energy of about 0.3 eV/atom, this reduces cluster fragmentation upon landing on the substrate so that the cluster-assembled film retains the memory of the structure and properties of the precursor clusters. In situ Raman measurements were performed using the 532 nm line of a frequency-doubled Nd:Yag and the 632.8 nm line of a He-Ne laser as excitation lights. Where not otherwise stated all the reported Raman spectra are acquired using the 532 nm line. The backscattered light was analyzed by a Jobin-Yvon T64000 spectrometer in triple grating configuration and detected by a liquid $N_2$-cooled CCD.

The temperature of the substrate can be varied in a range from 100 K to 350 K. When Raman spectra of films grown at a certain deposition temperature are reported, we mean that the sample was deposited and then kept at the same temperature during the Raman measurement, if not otherwise stated.

## III. Results

The in situ Raman spectra of cluster-assembled films deposited in ultra high vacuum conditions (~ $10^{-9}$ mbar) on substrates at 300 K, 160 K and 100 K respectively are reported in Fig 1. The spectra are characterized by the presence of G and D bands between 900 and 1700 cm$^{-1}$ typical of $sp^2$ amorphous carbon [43,44] and of a band between 1800 and 2200 cm$^{-1}$ (the so-called C band [20]) associated to stretching modes of $sp$ hybridized carbon. The C band is composed by a main peak at about 2100 cm$^{-1}$ (C2 peak) and a shoulder at a lower frequency (C1 peak at about 1980 cm$^{-1}$). Contributions from different electronic structures of carbon chains, i.e. polyynes (alternated single and triple bonds) for C2 peak and cumulenes (all double bonds) for C1 peak, have been proposed as responsible for the appearance of these features [19,20,24]. In the low-frequency part of the spectra two peaks at 450 cm$^{-1}$ and 720 cm$^{-1}$ are present.

The high-frequency Raman features are consistent with a disordered material where a high fraction of linear $sp$ carbon structures are embedded in a mainly $sp^2$ hybridized network [20,40]. The relative content of $sp$ structures can be evaluated, as proposed in previous works [19,24], by calculating the ratio $I_C^{rel} = I_C/I_{DG}$ between the integrated intensity of the C and D+G bands. In a similar way the relative intensity of cumulene and polyyne contributions can be evaluated through $I_{C1}^{rel}$ and $I_{C2}^{rel}$. By in situ near edge x-ray absorption fine structure (NEXAFS) characterization of cluster-assembled carbon films [45] an estimate of the relative Raman cross section of $sp$ structures (average value over the chain length distribution of both polyynes and cumulenes in the sample) has been recently performed on films deposited with the PMCS source, permitting to estimate a fraction of $sp$ bonds ranging from 10% to 25% in films deposited at room temperature [45].

A decrease in the deposition temperature, as shown in Fig.1, produces a substantial increment of the C band intensity, and thus of the $sp$ content, as shown by the values of the Raman relative intensity $I_C^{rel}$ reported in Table 1. The estimated fraction of $sp$ bonds in the films is also reported showing a $sp$ bond content reaching 50% (best estimate) for the film deposited at 100 K. Moreover, the C1 and C2 peak intensities reveal a strong increment of cumulene contribution with decreasing deposition temperature. As shown in Table 1 $I_{C1}^{rel}$ increases by more than 150% while $I_{C2}^{rel}$ increases by 45% going from 300 K down to 100 K deposition temperature.

The 450 cm$^{-1}$ and 720 cm$^{-1}$ features show a different behaviour with substrate temperature: the 450 cm$^{-1}$ peak becomes weaker with increasing deposition temperature and it is strongly reduced when the films are exposed to ambient atmosphere as observed for C band [19]. Conversely, the 720 cm$^{-1}$ feature does not show any appreciable evolution with temperature nor when the film is exposed to air as shown in Fig.2. It is noticeable that, after a prolonged exposition (several months) to the ambient, the 450 cm$^{-1}$ feature is almost completely absent as well as the C band while the 720 cm$^{-1}$ is still substantially unmodified. The same behaviour is observed when a film deposited at low temperature (160 K) is annealed in situ up to room temperature: a decrease of the C band and of the 450 cm$^{-1}$ feature is observed, whereas the 720 cm$^{-1}$ feature is not modified (see Fig.2).

In Fig.3 the relative intensity of the 450 cm$^{-1}$ and 720 cm$^{-1}$ peaks ($I_{450}^{rel} = I_{450}/I_{DG}$ and $I_{720}^{rel} = I_{720}/I_{DG}$) as a function of film temperature has been compared with the cumulene ($I_{C1}^{rel}$) and polyyne ($I_{C2}^{rel}$) behavior. $I_{450}^{rel}$ and $I_{C2}^{rel}$ (polyyne) show a parallel evolution, while $I_{720}^{rel}$ is substantially constant. Conversely the cumulene evolution ($I_{C1}^{rel}$) seems markedly different from both the low frequency bands.

Figure 4 reports Raman spectra of a film deposited and kept at 160 K for different excitation wavelengths (532 nm and 632.8 nm): by varying the excitation wavelength the D-G bands change in shape and position and the C band reveals strong modifications in the intensity and position of the cumulene and polyyne contribution as already discussed in a previous paper [24]. On the other hand the shape, the intensity and the position of the 450 cm$^{-1}$ and 720 cm$^{-1}$ peaks remain substantially unmodified showing no dependence on the Raman excitation wavelength.

### IV. Numerical Simulations

In order to analyze the low frequency Raman features in the 400-800 cm$^{-1}$ region, first principles calculations of bending modes in *sp* carbon chains have been carried out. Density functional theory (DFT) calculations of TO phonons of the infinite linear carbon chain have been recently published [5]. These data have been obtained through plane waves pseudopotential method as implemented in PWscf code [46] with the PBE exchange-correlation functional [47]. From these calculations we extracted the phonon density of states (PDOS) which is reported in Fig. 5. We also computed the off-resonance Raman responses due to low frequency bending modes of hydrogen capped carbon chains containing from N = 3 up to N = 5 triple bonds (i.e. 6-10 carbon atoms). The Gaussian03 package [48] has been used and the calculations employed the PBE functional and a polarized triple split valence basis set (cc-pVTZ).

The low frequency region of the simulated Raman spectra of hydrogen terminated chains is reported in Fig. 5. Different chain lengths from 6 to 10 carbon atoms have been considered not showing an appreciable dependence of the frequency of C−C≡C bending modes on the chain length. Both linear and bow-bent structures (bond angles range from 173 to 177 degrees) have been considered to take into account distorted configuration of *sp* chains in the *sp$^2$* amorphous network [41,42]. It has to be noticed that all the bent geometries are stable minima obtained from standard geometry optimization starting from a bent molecular shape. No significant modifications of the Raman bands is induced by a distortion of the chain configuration. The PDOS of the polyynic infinite chain in the 300-600 cm$^{-1}$ region shows a sharp peak at about 500 cm$^{-1}$ due to the TO phonon branch [5]. This peak in the PDOS nicely corresponds to the peak obtained from the calculation of the Raman response of finite chains. Also the vibrational displacements associated to the peak at 500 cm$^{-1}$ in finite chains correspond to the displacements associated to the TO phonon. No vibrational modes of polyynes are present in the 700-1000 cm$^{-1}$ frequency range.

The minor peaks in Fig.5 are not related to the TO bending phonon of the parent polyyne (i.e. the correspondent infinite chain). Instead, they are relative to the CH bending modes of the terminal bonds (at about 600 cm$^{-1}$) and to LA or TA phonon modes of the parent polyyne. In particular, these latter modes are dispersive with chain length due to the acoustic nature of the associated phonon (see Figure 5). It is worth to note that the CH bending modes are localized on the chain ends, while the vibrational displacements associated to the bending mode at about 500 cm$^{-1}$ do involve the carbon chain. Therefore it is possible to extrapolate the results on hydrogen capped chains to *sp* carbon chains embedded in a *sp$^2$* matrix, even if the details of the chemical bonding between *sp* carbon and the surrounding matrix are still unknown.

### V. Discussion

Low frequency Raman bands were observed in a wide variety of disordered carbon systems including diamond like carbon, carbon nitride, hydrogenated carbon nitride and carbon nano-onions [27-36]. In general, though disorder activated Raman modes corresponding to peaks in the vibrational density of states (VDOS) are invoked to justify the presence of such low Raman features, there is no complete agreement on the origin and assignment of these modes. In amorphous carbon 400 and 700 cm$^{-1}$ peaks were assigned to VDOS of both *sp$^2$* and mixed *sp$^2$-sp$^3$* [28, 29]. Zone centre phonons (M point) of out of plane transverse acoustic and transverse optic branch activated in curved graphene particles were invoked for the interpretation of peaks at 450 and 700 cm$^{-1}$, respectively, in the spectrum of carbon nano-onions and carbon nitride [34,35]. Moreover some authors observed the 700 cm$^{-1}$ peak only and assigned it to in-plane rotation of sixfold carbon rings (L mode) which corresponds to a peak in the phonon density of states PDOS of graphite [33,36]. Others assigned the same feature to *sp$^3$* bonding [32] or to VDOS of *sp$^3$-sp$^2$* mixed bonds [31].

Cluster-assembled carbon films deposited at low temperature are peculiar for the observed low frequency Raman features at about 450 and 720 cm$^{-1}$. Low frequency features have been already observed in *sp-sp$^2$* amorphous carbon in the past. Kuzmany and co-workers [49] observed a

420 cm$^{-1}$ peak in carbynoid structures stabilized by alkali metals. It was assigned by the authors to vibrational modes of the amorphous *sp*$^2$ carbon matrix in agreement with Li and Lannin [27] who discussed the occurrence of 450 cm$^{-1}$ and 750 cm$^{-1}$ Raman features in glassy carbon and in annealed amorphous carbon. Li and Lannin invoked a disorder induced Raman activation of peaks present in the vibrational density of states (VDOS) experimentally measured by neutron scattering.

In *sp-sp*$^2$ nanostructured carbon films the intensity of the 450 cm$^{-1}$ peak ($I_{450}^{rel}$) shows the same evolution of the polyyne contribution ($I_{C2}^{rel}$) as reported in Fig. 4 suggesting that this feature is produced by the *sp* fraction. Our numerical simulations indicate that a band around 450 cm$^{-1}$ is consistent with a convolution of bending modes of *sp* chains of different lengths. The frequency of such modes seems to be not strictly related with the chain length (i.e. number of atoms, conjugation length) or with the spatial arrangement of the chain (i.e. linear or bent conformation). This is also confirmed by the calculation of the modes of an infinite linear polyyne where a feature at about 500 cm$^{-1}$ characterizes the calculated PDOS. This occurrence helps us in this assignment since, even in the case of a very disordered material in which the actual structure and conformation of *sp* and *sp*$^2$ phase is substantially unknown, the expected Raman peak position should be substantially the same as in hydrogen terminated linear chains (our model system). Such extremely weak dependence on chain length and conformation may justify the observed behaviour of the 450 cm$^{-1}$ peak whose position does not change with varying the excitation wavelength. It has to be noticed that stretching modes of *sp* phase (the C band) behave in a complete different way showing a strong dependence on chain length and hence on excitation wavelength.

The reported observations and calculations indicate that the 450 cm$^{-1}$ band is directly related with bending modes of the *sp* phase. This does not exclude a possible minor contribution from the VDOS of the *sp*$^2$ network which also shows features falling in this region [27,50], as suggested by the observation of a remaining contribution after exposure to air. In principle, a contribution from cumulene structures cannot be a priori excluded since finite structures or highly distorted configurations could be responsible for the activation of Raman modes otherwise forbidden when dealing with perfect and infinite linear cumulene (i.e. acoustic modes only in one atom unit cell, see for instance [51]). It is reasonable in fact to assume the presence of curved or more complex configurations (even rings as suggested by Liu et al. [8]) of *sp* structures as outlined also by numerical simulations of metastable *sp-sp*$^2$ clusters [41,52].

Our experiments show that the 720 cm$^{-1}$ band is almost unchanged even as a function of excitation light suggesting a possible attribution to a feature in the VDOS (according to the Shuker and Gammon relation) of the *sp*$^2$ network which is Raman activated by the disorder of the amorphous structure, in agreement with Li and Lannin. The conditions for selection rule relaxation and for 720 cm$^{-1}$ peak activation are always satisfied since even after the strong structural modifications produced by a prolonged exposure to the ambient, the film is still amorphous, as attested by the shape of the G and D bands.

### VI. Conclusions

We have characterized by in situ Raman spectroscopy nanostructured *sp-sp*$^2$ carbon films with coexisting cumulene-like, polyyne-like and *sp*$^2$ structures obtained by depositing a supersonic beam of carbon clusters with *sp* and *sp*$^2$ hybridization [41,52]. Two bands at 450 cm$^{-1}$ and 720 cm$^{-1}$ have been detected. Both the C-C stretching at 1800-2200 cm$^{-1}$ and the 450 cm$^{-1}$ features decrease showing a parallel evolution upon thermal annealing or exposure to the ambient atmosphere. The 720 cm$^{-1}$ peak does not show this kind of dependence. Calculations of the vibrational modes of finite and infinite *sp* structures (polyynes) show active Raman modes and a peak in the PDOS at about 500 cm$^{-1}$ while no modes are predicted in the 700-1000 cm$^{-1}$ region. On this basis we can attribute the 450 cm$^{-1}$ peak to bending modes of *sp* structures with polyyne-like structure and the 720 cm$^{-1}$ band to a peak in the VDOS of the *sp*$^2$ amorphous phase activated by disorder since it does not show any substantial modifications even when changing the excitation wavelength.

These results reveal that, in addition to the stretching C-C modes above 1800 cm$^{-1}$, new features related with bending modes are present in the Raman spectra of this novel form of solid carbon consisting in an amorphous *sp-sp$^2$* network.

**Acknowledgment**

The authors affiliated at the University of Milano acknowledge financial support from Project PRIN "Novel approach to growth and characterization of carbon-based nanostructured and nanocomposite materials with extended interface". The authors affiliated at Politecnico di Milano acknowledge financial support from FlagProject *"ProLife mobilità sostenibile"* funded by the Milano city administration.

# Table Caption

Table 1: *sp* phase (polyyne and cumulene) contribution for films deposited at different temperatures. The estimated *sp* bond fraction on the basis of NEXAFS results (see the text) is also reported

# Figure Captions

Figure 1: Raman spectra of films deposited and maintained at 100, 160 and 300 K in UHV conditions, respectively. Magnification of the low frequency region is reported in the left panel with Lorentzian fits.

Figure 2: Raman spectra in the low frequency region (100-900 cm-1) of a film deposited at 160 K, in situ annealed up to 300 K, immediately after exposure to ambient conditions and after prolonged exposure. The D-G and C bands of the same films are shown in the inset.

Figure 3: Temperature evolution of the relative intensity of cumulene ($I_{C1}^{rel}$), polyyne ($I_{C2}^{rel}$), 450 ($I_{450}^{rel}$) and 720 ($I_{720}^{rel}$) cm$^{-1}$ Raman bands. The intensity behavior of the 450 cm$^{-1}$ and 720 cm$^{-1}$ peaks have been evaluated by a lorentian fit (Fig.1) and then normalized to the total intensity of the D-G bands in order to be directly compared to the intensity behavior of the C band.

Figure 4: Raman spectra at different excitation wavelengths (633 and 532 nm) of films deposited at 160 K.

Figure 5: Comparison between the TO phonon DOS and the low frequency Raman active modes of hydrogen capped polyynes with N = 3 up to N = 5 triple bonds (i.e. 6-10 carbon atoms). The left panel reports data computed on perfectly linear molecules, while the right panel report results from bow-bent molecules. The strongest Raman lines in this frequency range refer to bending modes with nuclear displacements related to the TO phonon of the infinite chain. The minor lines relative to CH bending modes, TA and LA modes are marked . The other unmarked lines (for N = 4 and N = 5) are still TA modes with an increasing number of nodes. Bottom: nuclear displacements associated with the strongest Raman lines reported in the simulated spectra.

| Deposition temperature T(K) | Cumulenic contribution $I_{C1}$ | Polyynic contribution $I_{C2}$ | Total sp phase contribution $I_C$ | Estimated sp bond fraction(%) |
|---|---|---|---|---|
| 100 K | 0.47 | 0.13 | 0.60 | 20-50% |
| 160 K | 0.40 | 0.11 | 0.51 | 17-42% |
| 300 K | 0.17 | 0.09 | 0.26 | 10-25% |

Table 1

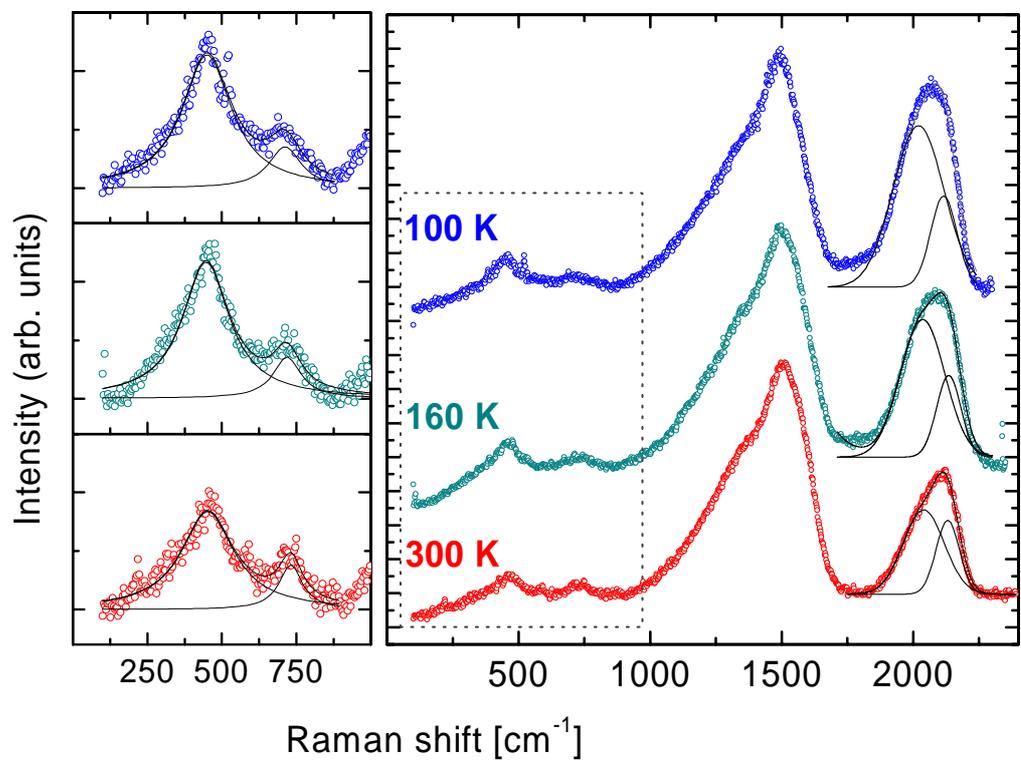

Figure 1

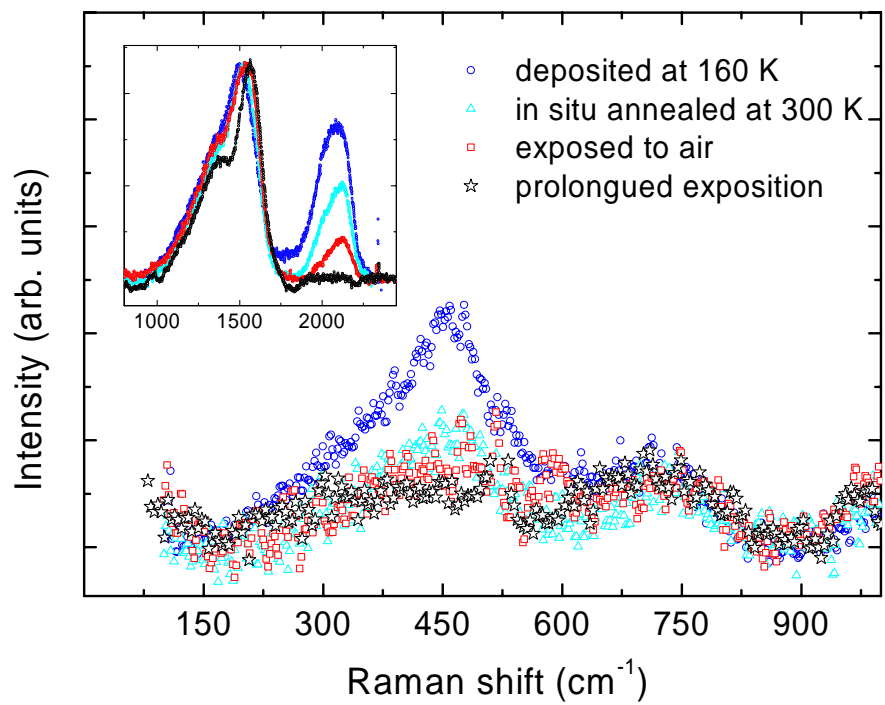

Figure 2

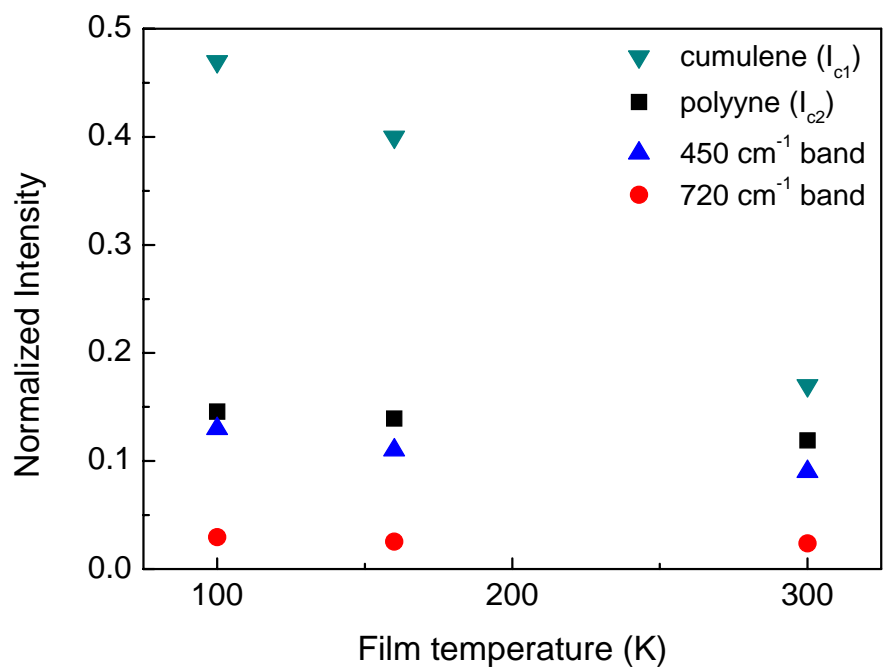

Figure 3

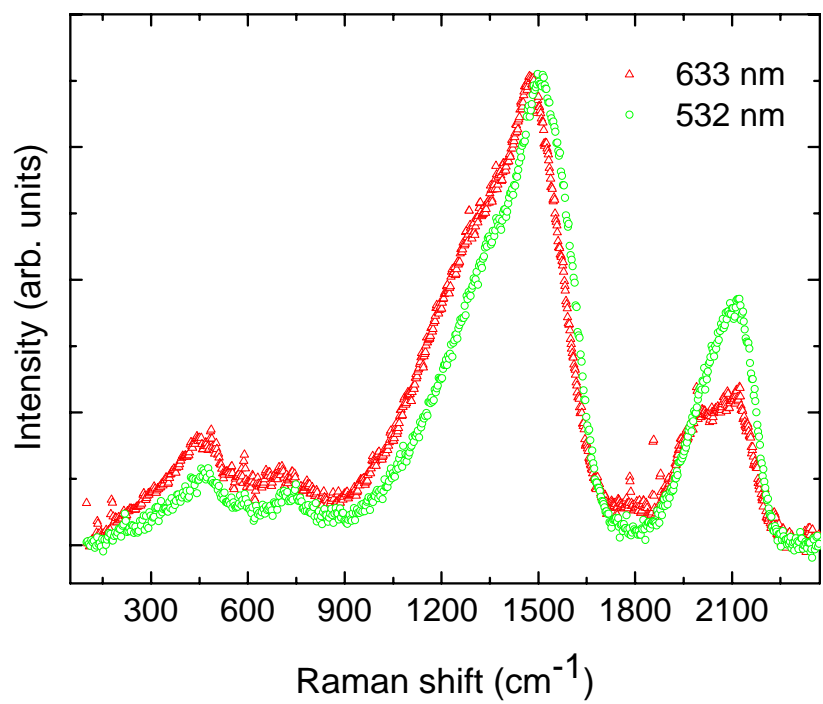

Figure 4

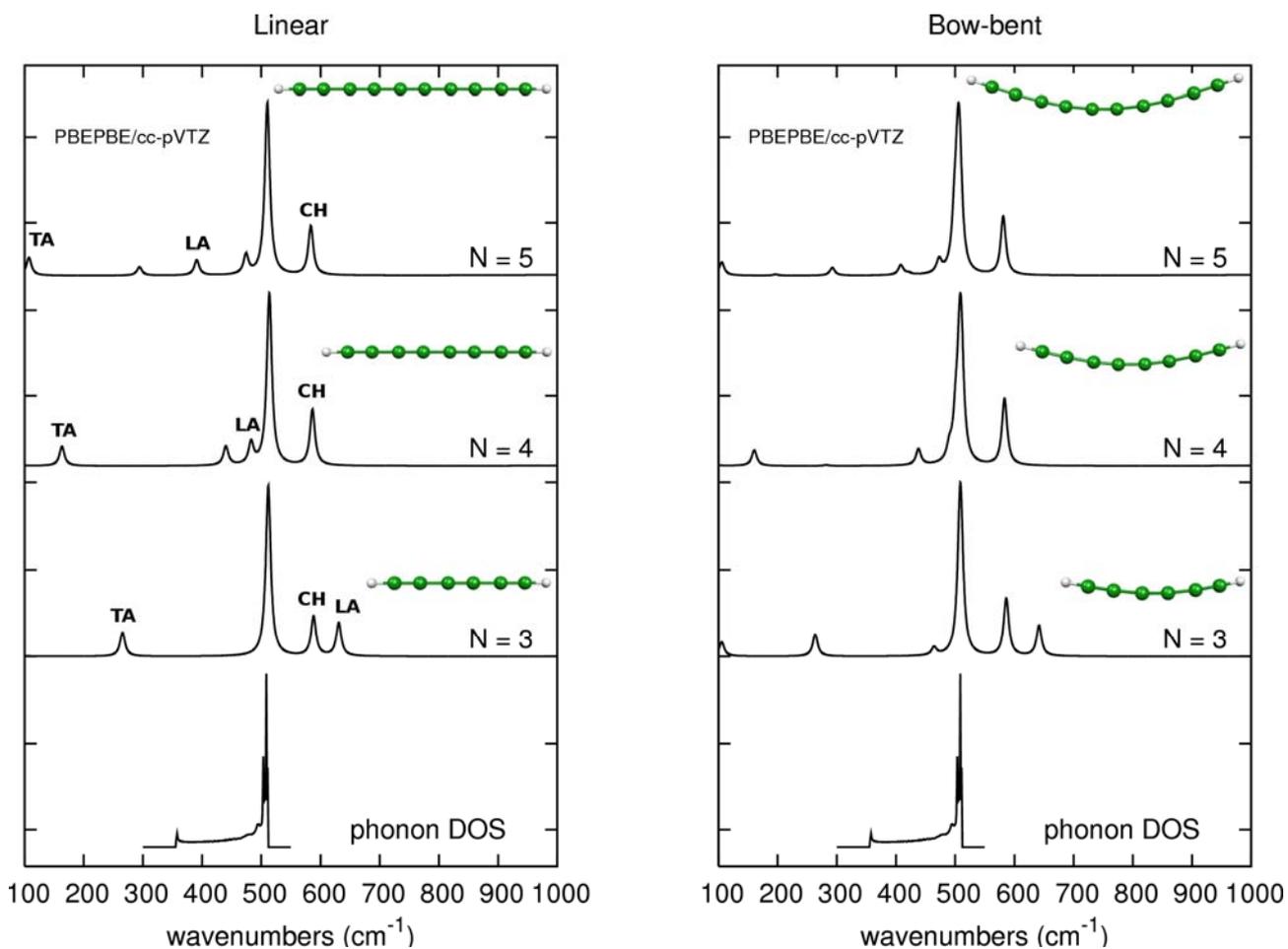

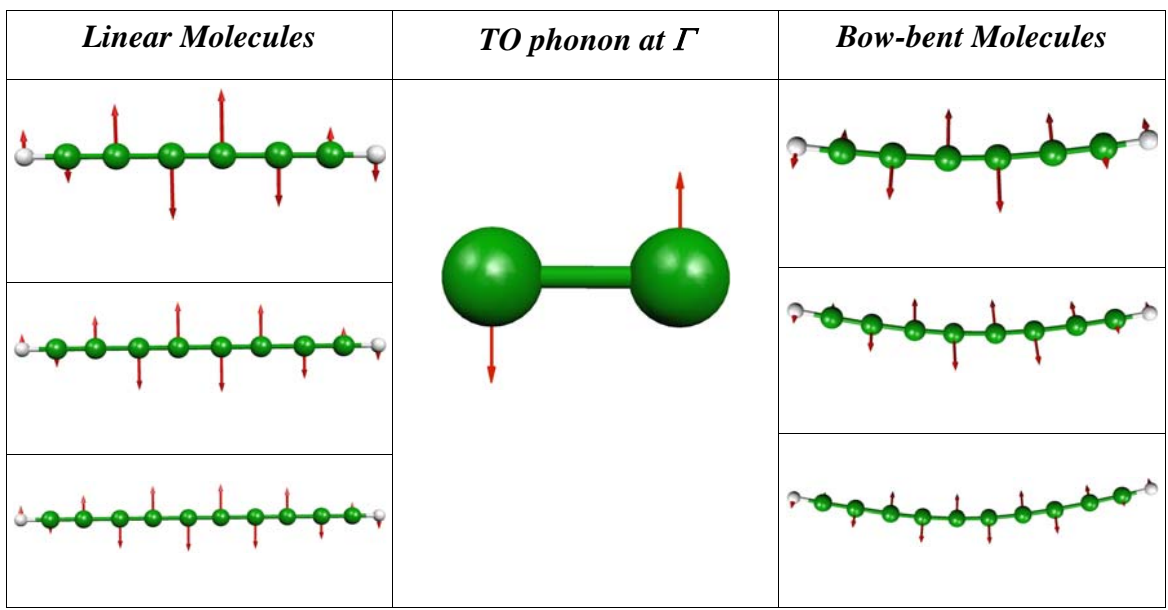

Figure 5